\documentclass[12pt]{iopart}

\usepackage{amsmath,amsthm,amsfonts,amssymb,bm,epsfig}

\def\mathbi#1{\textbf{\emph #1}}
\DeclareRobustCommand\openone{\leavevmode\hbox{\small1\normalsize\kern-.33em1}}

\begin{document}

\title{Encoding a qubit into multilevel subspaces}

\author{Matthew Grace$^{1}$, Constantin Brif$^{1}$, Herschel
  Rabitz$^{1}$, Ian Walmsley$^{2}$, Robert Kosut$^{3}$, and Daniel
  Lidar$^{4}$\footnote{Current address: Chemistry and Electrical
  Engineering Departments, University of Southern California, Los
  Angeles, CA 90089}}

\address{$^{1}$ Department of Chemistry, Princeton University,
Princeton, New Jersey 08544}

\address{$^{2}$ Department of Physics, University of Oxford, Oxford OX1
3PU, UK}

\address{$^{3}$ SC Solutions, Inc., 1261 Oakmead Parkway, Sunnyvale, CA
94085}

\address{$^{4}$ Department of Chemistry, University of Toronto, Toronto,
Ontario, M5S 3H6, Canada}

\eads{\mailto{mgrace@princeton.edu}, \mailto{cbrif@princeton.edu},
  \mailto{hrabitz@princeton.edu}, \mailto{walmsley@physics.ox.ac.uk},
  \mailto{kosut@scsolutions.com}, and \mailto{lidar@usc.edu}}



\begin{abstract}
We present a formalism for encoding the logical basis of a qubit into
subspaces of multiple physical levels.  The need for this multilevel
encoding arises naturally in situations where the speed of quantum
operations exceeds the limits imposed by the addressability of
individual energy levels of the qubit physical system.  A basic feature
of the multilevel encoding formalism is the logical equivalence of
different physical states and correspondingly, of different physical
transformations.  This logical equivalence is a source of a significant
flexibility in designing logical operations, while the multilevel
structure inherently accommodates fast and intense broadband controls
thereby facilitating faster quantum operations.  Another important
practical advantage of multilevel encoding is the ability to maintain
full quantum-computational fidelity in the presence of mixing and
decoherence within encoding subspaces.  The formalism is developed in
detail for single-qubit operations and generalized for multiple qubits.
As an illustrative example, we perform a simulation of closed-loop
optimal control of single-qubit operations for a model multilevel
system, and subsequently apply these operations at finite temperatures
to investigate the effect of decoherence on operational fidelity.
\end{abstract}


\submitto{\NJP}

\maketitle

\section{Introduction}

Quantum computation is an extremely active area of research, with
potential realizations ranging from solid-state to atomic and optical
systems \cite{NiCh00, QIST}.  Following seminal work on ideal quantum
computation, the focus of research has shifted in recent years to
practical issues of implementing scalable quantum gates in real systems
in the presence of noise, decoherence, and imperfect controls.  One
significant issue is the fundamental process of decoherence caused by
coupling to the environment.  Another obstacle is noise that, in any
practical situation, exists in external control fields which are
normally required to perform the logical operations of quantum
information processing \cite{EnKi01}.  In order to ensure fault-tolerant
operation of quantum registers, errors, including those associated with
environmentally-induced decoherence and control field noise, must be
minimized below a specified threshold \cite{Shor96}.  Since external
controls are essential for logical operations, it is reasonable to
assume that practical implementations of quantum computation, especially
in relatively complex physical systems, can be greatly facilitated by
the powerful methods of optimal control \cite{RVM00, RaWa03}.
Closed-loop adaptive laboratory control is of special importance because
of its inherent robustness to noise and suitability for managing complex
physical systems \cite{RVM00, RaWa03, JuRa92, ZhRa94, DeRa98, GZR00,
TuRa01, HMV02}.  Furthermore, intense ultrafast control fields, which
are essential for accelerating quantum operations, thereby diminishing
the effect of decoherence, are especially suited for closed-loop
laboratory control.  Applying the methods of optimal control to various
realizations of quantum computation was recently considered in a number
of works \cite{RaRa96, SKH99, BRWW01, TeVR01, TeVR02, TeVR03, PaKo02,
Kosl02a, Kosl02b, SkTa04}.

The majority of schemes proposed for physical realizations of quantum
computation assign a single physical level to represent each of the
logical basis states ($|0\rangle$ and $|1\rangle$) of a qubit.  In such
a situation, each logical operation is represented by a unique physical
transformation \cite{NiCh00}.  In this paper we consider a more general
situation, where each logical basis state is encoded into a subspace of
multiple physical levels.  This generalization leads to the logical
equivalence of different physical states, i.e., different multilevel
superpositions can represent the same logical state of a qubit.
Correspondingly, with multilevel encoding (MLE) there will exist
different physical transformations which will also be logically
equivalent, i.e., these transformations represent the same logical
operation.

An important question is whether the use of MLE yields any practical
advantages in comparison to the usual single-level encoding (SLE).  The
following analysis indicates that MLE is a natural choice when the need
to accelerate quantum operations requires the bandwidth of control
fields to exceed the separation between energy levels of a quantum
register.  For systems with a dense energy spectrum, the factor
determining individual level addressability is the ratio between the
bandwidth, $\Delta\Omega$, of the external control field and the
characteristic energy-level spacing, $\Delta\omega$, of the physical
system that realizes the register.  In the limit of slow control, i.e.,
$\Delta\Omega < \Delta\omega$, logical basis states can be encoded into
single physical levels and MLE is not necessary.  While such slow
controls may be satisfactory for proof-of-principle experiments,
practical realizations of quantum registers in many cases will require
much faster operations and correspondingly, faster controls, with
$\Delta\Omega \gg \Delta\omega$.  In such a case, one must take into
account the various transitions between the multiple states driven by a
control field.  Encoding each logical basis state into a subspace of
multiple physical levels circumvents this issue, since addressability of
individual levels within each subspace is not necessary due to the
principle of logical equivalence developed in this work.

It is clear from the arguments above that the importance of MLE depends
on the specific physical realization of a quantum register, so it will
be helpful to consider specific examples of systems which may be
suitable for realizations of MLE.  One such system is a molecule in
which the logical basis of a qubit is realized by vibrational states on
the ground and/or excited electronic surfaces \cite{ZKLA01, BGLA02}.
For example, typical vibrational periods in homonuclear alkaline dimers
range from 300 to 500 fs \cite{WaWa98}.  If the system is driven by
ultrafast laser pulses of durations ranging from 20 to 100 fs (which are
currently successfully used in closed-loop adaptive control experiments
with molecular systems \cite{Weiner00, Gerber01, Gerber03, Levis01,
Levis02, Silberberg02, Silberberg03}), many transitions between
vibrational levels will be driven simultaneously.  In such a case, it
can be useful to encode each logical basis state into a set of multiple
molecular levels.

Another interesting example is a system whose internal levels are used
for the logical basis while it is spatially confined in an external
potential (e.g., an ion in an electromagnetic trap or an atom in an
optical lattice).  In the majority of schemes proposed for quantum
computation with trapped ions, the gate time is limited by the trap
frequency, mainly due to the need to spectroscopically resolve the
motional sidebands \cite{CiZo95, PCZ98, SoMo, JPK00, MSJ00}.  This is a
serious limitation because typical ion-trap frequencies are relatively
low, ranging from 100~kHz to 100~MHz \cite{DJW98, CMDW03}.  External
controls which are faster than the vibrational frequency of a trapped
qubit will inevitably excite multiple motional levels.  This effect will
be especially important outside of the Lamb-Dicke limit, i.e., when
multiple sidebands are excited by a laser field interacting with a
trapped ion.  Encoding each logical basis state into multiple motional
levels of a trapped particle can be a useful approach in situations when
fast controls are desired.

An idea related to the concept of MLE was recently proposed by
Garc\'{i}a-Ripoll \emph{et al.} \cite{GZC03}, who considered a two-qubit
operation acting on a pair of trapped ions.  The essence of this
proposal is to apply coherent laser controls which leave the initial
motional state unchanged at the end of the sequence of control pulses,
thereby allowing for operations which are much faster than the trap
frequency.  MLE allows for more general operations (not just the
identity operation) on the motional levels which constitute the encoding
subspaces for a trapped particle.  However, MLE may not be directly
applicable to popular schemes where the common motional mode of a number
of trapped ions is additionally employed to entangle the internal atomic
states of different qubits.  At this point it is not yet clear which
systems will be most suitable for MLE in practical realizations of
quantum computation, but the importance of MLE will increase in line
with the acceleration of quantum operations.

Searching for an optimal control field to perform a logical operation on
a multilevel system can benefit from MLE due to the logical equivalence
of different physical transformations.  Also, the multiplicity of
logically equivalent physical transformations can facilitate finding
optimal operations with improved robustness against noise in external
controls \cite{BRMR05}.  Moreover, multiple physical transformations
which realize the same logical operation may be used interchangeably
throughout the computation, providing greater flexibility in the
laboratory implementation of controls.

Perhaps the most important property of MLE is its utility as a practical
instrument to reduce the effects of mixing in the initial quantum state
(e.g., due to thermal excitations) and decoherence throughout the
computation.  In many practical situations, the initialization of a
quantum register requires cooling a system to the ground state since the
impurity of the initial state due to thermal excitations reduces
quantum-computational fidelity.  This cooling is often a difficult and
slow process, therefore a practical alternative is desirable.  Another
significant source of errors is decoherence during logical operations.
A possible way to overcome these problems is, once again, by encoding
the logical basis into subspaces of multiple physical levels.  Due to
the principle of logical equivalence, mixing and decoherence
\emph{within} each encoding subspace do not affect coherence
\emph{between} the two logical basis states and therefore do not reduce
computational fidelity.

This paper is organized as follows.  In section~\ref{sec:concept}, the
concept and mathematical formalism of multilevel encoding are developed
for a single qubit.  Logical equivalence resulting from MLE is defined
for physical states and transformations, and equivalence classes of
logical states and operations are introduced.  In
section~\ref{sec:1-qubit}, the MLE formalism and corresponding logical
equivalence are used to construct single-qubit unitary operations and
derive their general form (a tensor-product structure).
Section~\ref{sec:dense} extends the MLE formalism to mixed states and
non-unitary operations.  In particular, this allows for logical
equivalence between pure and mixed states, and between unitary and
general non-unitary (permitting decoherence) quantum operations,
provided that full coherence exists between the encoding subspaces.  In
section~\ref{sec:example}, we demonstrate the formalism and some of its
advantages by performing a numerical simulation of closed-loop optimal
control of single-qubit operations for a model system based on the
electronic and vibrational levels of an alkaline dimer, and applying the
resulting operations at finite temperatures. In
sections~\ref{sec:2-qubits} and \ref{sec:m-qubits}, the MLE formalism is
generalized for logical states and operations of multiple qubits.  As in
the single-qubit case, a general tensor-product form for multi-qubit
unitary operations is obtained. Section~\ref{sec:conc} concludes the
paper with a brief summary of the results, open questions, and future
directions.

\section{Concept and Formalism}
\label{sec:concept}

\begin{figure}
\epsfxsize=0.35\textwidth \centerline{\epsffile{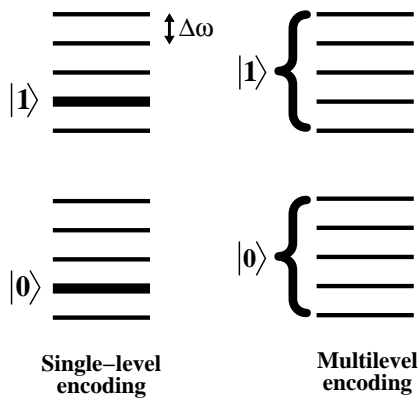}}
\caption{A sample scheme that illustrates the difference between
  single-level and multilevel encoding of logical basis states of a
  qubit.}
\label{Fig1}
\end{figure}

For a single qubit, MLE involves partitioning the energy levels of a
given quantum system into two distinct subspaces of equal dimension, the
``encoding subspaces'' $\mathcal{S}_{0}$ and $\mathcal{S}_{1}$.  Let
$|\mathbi{a}\rangle$ denote \emph{any} element of $\mathcal{S}_{0}$ such
that $\langle\mathbi{a}|\mathbi{a}\rangle = 1$ and $|\mathbi{b}\rangle$
denote \emph{any} element of $\mathcal{S}_{1}$ such that
$\langle\mathbi{b}|\mathbi{b}\rangle = 1$, i.e.,
\begin{gather}
\label{encode_a}
\nonumber |\mathbi{a}\rangle = \sum_{i = 1}^{n} a_{i} |\chi_{i}\rangle =
\Big( a_{1} ~\cdots~ a_{n} \Big)^{T} \\ \text{such that} \
\langle\mathbi{a}|\mathbi{a}\rangle = \sum_{i = 1}^{n} |a_{i}|^{2} = 1
\end{gather}
and
\begin{gather}
\label{encode_b}
 \nonumber |\mathbi{b}\rangle = \sum_{i = 1}^{n} b_{i} |\phi_{i}\rangle
= \Big( b_{1} ~\cdots~ b_{n} \Big)^{T} \\ \text{such that} \
\langle\mathbi{b}|\mathbi{b}\rangle = \sum_{i = 1}^{n} |b_{i}|^{2} = 1.
\end{gather}
Here $n$ is the dimension of the encoding subspaces and $\{
|\chi_{i}\rangle \}$ and $\{|\phi_{i}\rangle \}$ are orthonormal bases
that span $\mathcal{S}_{0}$ and $\mathcal{S}_{1}$, respectively.
Correspondingly, $\langle\chi_{i}|\chi_{j}\rangle =
\langle\phi_{i}|\phi_{j}\rangle = \delta_{i j}$ and
$\langle\chi_{i}|\phi_{j}\rangle = 0$.  It is important to emphasize
that for each subspace ($\mathcal{S}_{0}$ or $\mathcal{S}_{1}$) there
exists an infinite set of states that satisfy the respective criteria of
(\ref{encode_a}) or (\ref{encode_b}).  Within this framework each
logical basis state is encoded into one of these subspaces (figure
\ref{Fig1}).  Therefore, the Hilbert space of an encoded qubit (which is
two-dimensional in the SLE representation) corresponds to an expanded
(and higher-dimensional) Hilbert space of physical levels.  In this
expanded Hilbert space the logical basis states are expressed as
\begin{subequations}
\label{def}
\begin{gather}
|0\rangle = |0\rangle_{\text{L}} \otimes |\mathbi{a}\rangle = \Big(
 a_{1} ~\cdots~ a_{n} ~\underbrace{0 ~\cdots~ 0}_{n} \Big)^{T} \\
 |1\rangle = |1\rangle_{\text{L}} \otimes |\mathbi{b}\rangle = \Big(
 \underbrace{0 ~\cdots~ 0}_{n} ~b_{1} ~\cdots~ b_{n} \Big)^{T},
\end{gather}
\end{subequations}
where $|0\rangle_{\text{L}}=(1,0)^{T}$ and
$|1\rangle_{\text{L}}=(0,1)^{T}$ are the ``logical components'' of the
logical basis states and $|\mathbi{a}\rangle$ and $|\mathbi{b}\rangle$
are the ``encoded components.''  It follows that $\langle 0|0\rangle =
\langle 1|1\rangle = 1$ and $\langle 0|1\rangle = 0$.  Thus, any
superposition state $|\psi\rangle$ can be written as
\begin{align}
\label{qubit}
\nonumber |\psi\rangle & = c_{0} |0\rangle + c_{1} |1\rangle \\ & =
c_{0}|0\rangle_{\text{L}} \otimes |\mathbi{a}\rangle +
c_{1}|1\rangle_{\text{L}} \otimes |\mathbi{b}\rangle,
\end{align}
where $|\psi\rangle$ is a vector of dimension $2n$ and $| c_{0} |^{2} +
| c_{1} |^{2} = 1$ so that $\langle\psi|\psi\rangle = 1$.  Throughout
this work we employ the following abbreviated notation:
\begin{equation}
|\psi\rangle = \left( \begin{array}{c} \alpha \\ \beta
\end{array} \right),
\end{equation}
where $\alpha = c_{0} |\mathbi{a}\rangle$ and  $\beta = c_{1}
|\mathbi{b}\rangle$.

There are many different physical situations in which a tensor-product
structure (TPS) in the form of (\ref{qubit}) can be realized
\cite{ZLL04}.  An important point is that in the case of MLE this TPS is
\emph{virtual} since it does not arise from spatially separated degrees
of freedom, instead it is an abstract partitioning of the Hilbert space.
This essential feature makes MLE physically quite different from the
mathematically similar structure of quantum error correction (QEC)
\cite{KnLa97}.  In the case of QEC one deals with a number of distinct
physical systems (a qubit and ancillae) which become entangled via a set
of controlled interactions between them.  However, in the context of MLE
we deal with a single quantum system for each qubit (even if this system
is in fact composed of many particles, e.g., a molecule), and use of the
tensor product is a convenient mathematical description.  The Hilbert
space of a molecule or atom is just one example of a state space in
which a virtual TPS can be built.  Other, very general mechanisms of
producing a TPS come from superselection rules and the representation
theory of algebras of observables \cite{ZLL04}.  Therefore, there is a
plethora of possibilities for building an MLE structure: from single
atoms to molecules, to complex multiparticle systems.

As seen from (\ref{def}) and (\ref{qubit}), different physical states,
i.e., different choices of $|\mathbi{a}\rangle$ and
$|\mathbi{b}\rangle$, can represent the same logical state.  In order to
rigorously formulate this principle of logical equivalence, we introduce
a set of ``equivalence assay'' (EA) operators:
\begin{subequations}
\begin{align}
& \Lambda_{x} = \frac{1}{2} \sum_{i = 1}^{n}
\left(|\chi_{i}\rangle\langle\phi_{i}| +
|\phi_{i}\rangle\langle\chi_{i}|\right), \\ & \Lambda_{y} = \frac{1}{2
i} \sum_{i = 1}^{n} \left(|\chi_{i}\rangle\langle\phi_{i}| -
|\phi_{i}\rangle\langle\chi_{i}|\right), \\ & \Lambda_{z} =  \frac{1}{2}
\sum_{i = 1}^{n} \left( |\chi_{i}\rangle\langle\chi_{i}| -
|\phi_{i}\rangle\langle\phi_{i}| \right).
\end{align}
\end{subequations}
The EA operators can also be rewritten in the matrix form
\begin{equation}
\label{EA-ops}
\Lambda_{r} = \lambda_{r} \otimes \openone_{n} = \frac{1}{2} \sigma_{r}
\otimes \openone_{n}, \ \ r = \{ x, y, z \},
\end{equation}
where $\{ \sigma_{x}, \sigma_{y}, \sigma_{z} \}$ are the three Pauli
matrices, $\{ \lambda_{x}, \lambda_{y}, \lambda_{z} \}$ are the $2
\times 2$ matrices ($\lambda_{r} = \frac{1}{2} \sigma_{r}$) which
represent the three generators of the SU(2) group, and $\openone_{n}$ is
the $n \times n$ identity matrix.  Expectation values of the EA
operators of an arbitrary state $|\psi\rangle$ are
\begin{subequations}
\begin{align}
& \langle\psi|\Lambda_{x}|\psi\rangle =  \text{Re} \Big( c_{0}^{*} c_{1}
 \langle\mathbi{a}|\mathbi{b}\rangle \Big) = \text{Re}\left(
 \alpha^{\dagger} \beta \right), \\ &
 \langle\psi|\Lambda_{y}|\psi\rangle = \text{Im} \Big( c_{0}^{*} c_{1}
 \langle\mathbi{a}|\mathbi{b}\rangle \Big) = \text{Im} \left(
 \alpha^{\dagger} \beta \right), \\ &
 \langle\psi|\Lambda_{z}|\psi\rangle = \frac{1}{2} \Big( | c_{0} |^{2}
 \langle\mathbi{a}|\mathbi{a}\rangle - | c_{1} |^{2}
 \langle\mathbi{b}|\mathbi{b}\rangle \Big) = \frac{1}{2} \left( | c_{0}
 |^{2} - | c_{1} |^{2} \right).
\end{align}
\end{subequations}
Two physical states, $|\psi\rangle$ and $|\psi'\rangle$, are logically
equivalent, which is denoted as $|\psi\rangle \sim |\psi'\rangle$, if
they satisfy
\begin{equation}
\label{log-equiv}
\langle\psi| \Lambda_{r} |\psi\rangle = \langle \psi' | \Lambda_{r} |
\psi'\rangle, \ \  r = \{ x,y,z \}.
\end{equation}
In addition, the states must be normalized $( \langle\psi|\psi\rangle =
\langle\psi| \openone_{2} \otimes \openone_{n} |\psi\rangle = 1$ for all
$|\psi\rangle)$.  Logical equivalence resulting from MLE is analogous to
equality in the context of SLE.  Since the encoded component of all EA
operators is the identity operator $\openone_{n}$, logical equivalence,
as defined by (\ref{log-equiv}), allows for significant freedom in the
choice of the encoded components of a state without making it logically
different.  On the other hand, the logical components of the EA
operators, $\sigma_{r}$, together with the identity operator
$\openone_{2}$, span the space of all $2 \times 2$ matrices, ensuring
that all equivalent states are logically the same, i.e, that no logical
operation can distinguish between them.

Logical equivalence is a transitive property, i.e., if $|\psi\rangle
\sim |\psi'\rangle$ and $|\psi\rangle \sim |\psi''\rangle$, then
$|\psi'\rangle \sim |\psi''\rangle$.  Therefore, all states which are
mutually equivalent form an equivalence class $\mathcal{E}_{\psi}$,
which is formally expressed as
\begin{equation}
\mathcal{E}_{\psi} = \{ |\psi_{s}\rangle : |\psi_{s}\rangle \sim
|\psi\rangle \}.
\end{equation}
Correspondingly, every physical state in a given equivalence class
represents the \emph{same} logical state.

Logical equivalence can also be extended to quantum operations.  Two
operations, $U$ and $U'$ are logically equivalent (denoted as $U \sim
U'$), if
\begin{equation}
U |\psi\rangle \sim U' |\psi\rangle \ \ \text{for any} \ |\psi\rangle.
\end{equation}
Logical equivalence for operations is also a transitive property, i.e.,
if $U \sim U'$ and $U \sim U''$, then $U' \sim U''$.  Therefore, all
physical transformations which are mutually equivalent form an
equivalence class and every transformation in that class represents the
\emph{same} logical operation.

\section{Single-Qubit Operations for Multilevel Encoding}
\label{sec:1-qubit}

Any $2 \times 2$ unitary matrix can be represented as a linear
combination of the identity matrix $\openone_{2}$ and the three Pauli
matrices $\{ \sigma_{x}, \sigma_{y}, \sigma_{z} \}$.  The Pauli matrices
(up to a numerical factor)  are the generators of the Lie group SU(2)
and satisfy a set of commutation and multiplication
properties. Single-qubit logical operations for SLE are represented by
elements of the U(2) group.  In this section we develop the
corresponding representation of single-qubit logical operations within
the context of MLE.

Using the partitioning scheme for logical basis states described in
section~\ref{sec:concept}, the general form of a unitary operation
acting on a qubit encoded in the expanded ($2n$)-dimensional space is
\begin{equation}
\label{U-mas}
U = \left( \begin{array}{cc} A & B \\ C & D
\end{array} \right),
\end{equation}
where each sub-block $A$, $B$, $C$, and $D$ is an $n \times n$ matrix
representing operations on and between the two encoding subspaces,
$\mathcal{S}_{0}$ and $\mathcal{S}_{1}$.  Note that any operation $U$ is
defined up to an overall phase factor.

\subsection{The weak identity operation}

The \emph{weak} identity operation, $U_{I}$, is defined by the following
property: under the conditions of logical equivalence, $U_{I}$ leaves
the equivalence class $\mathcal{E}_{\psi}$ invariant, i.e., $U_{I}
\mathcal{E}_{\psi} = \mathcal{E}_{\psi}$, which can be expressed as
\begin{equation}
U_{I}|\psi\rangle \sim |\psi\rangle  \ \ \text{for any} \ |\psi\rangle.
\end{equation}
Using the general decomposition of (\ref{U-mas}), the equation above is
recast in the form
\begin{equation}
\left( \begin{array}{c} A \alpha + B \beta \\ C \alpha + D \beta
\end{array} \right) \sim
\left( \begin{array}{c} \alpha \\ \beta
\end{array} \right).
\end{equation}
We find the explicit form of $U_{I}$ by using the conditions of logical
equivalence, (\ref{log-equiv}).  The condition for $\Lambda_{z}$,
combined with the normalization property, requires that
\begin{subequations}
\begin{gather}
\left( \alpha^{\dagger} A^{\dagger} + \beta^{\dagger}B^{\dagger} \right)
\left( A \alpha + B \beta \right) = \alpha^{\dagger} \alpha , \\ \left(
\alpha^{\dagger} C^{\dagger} + \beta^{\dagger} D^{\dagger} \right)
\left(C \alpha + D \beta \right) = \beta^{\dagger} \beta,
\end{gather}
\end{subequations}
which is true if $A$ and $D$ are unitary and $B = C = 0$.  Using these
results, we next consider the two other conditions for equivalence
($\Lambda_{x}$ and $\Lambda_{y}$):
\begin{equation}
\alpha^{\dagger} A^{\dagger} D \beta \pm \beta^{\dagger} D^{\dagger} A
\alpha  = \alpha^{\dagger} \beta \pm \beta^{\dagger} \alpha,
\end{equation}
which are satisfied if $A = D$.  Therefore, we obtain
\begin{equation}
\label{UI}
U_{I} = \left( \begin{array}{cc} V_{0} & 0 \\ 0 & V_{0}
\end{array} \right) =
\openone_{2} \otimes V_{0},
\end{equation}
where $V_{0}$ is an arbitrary $n \times n$ unitary matrix and $0$
denotes the $n \times n$ matrix of zeros.  Therefore, the weak identity
$U_{I}$ is in fact a class of logically equivalent operations:
\begin{equation}
U_{I} = \{ \openone_{2} \otimes V_{0} : V_{0} \in \text{U}(n) \}.
\end{equation}
The identity operation $\openone_{2n}$ is also a member of this
equivalence class (corresponding to $V_{0} = \openone_{n}$).

The notion of the weak identity operation clarifies the meaning of
logical equivalence between different physical states.  For any state
$|\psi\rangle = (\alpha, \beta )^{T}$, the equivalence class
$\mathcal{E}_{\psi}$ is generated by the set of unitary operations
$U_{I}$ acting on $|\psi\rangle$:
\begin{equation}
\label{Epsi-class}
\mathcal{E}_{\psi} = \{ ( \openone_{2} \otimes V_{0} ) |\psi\rangle :
V_{0} \in \text{U}(n) \}.
\end{equation}
It is impossible to preserve logical equivalence when different
transformations act on $\alpha$ and $\beta$, e.g., the state
$|\tilde{\psi}\rangle = (V \alpha, V' \beta )^{T}$ with $V'  \neq V$ is
logically different from the state $|\psi\rangle = ( \alpha, \beta
)^{T}$ since the relative phase between $|0\rangle$ and $|1\rangle$ is
changed. This is why the tensor-product structure of (\ref{UI}) is
necessary for maintaining logical equivalence and emphasizes why logical
equivalence is defined for a superposition state $|\psi\rangle$.  It
would be meaningless to define logical equivalence for the individual
logical basis states since no definite relative phase between
$|0\rangle$ and $|1\rangle$ would exist.

It follows from (\ref{Epsi-class}) that an infinite number of different
physical states within an equivalence class correspond to the infinite
number of different choices for the unitary transformation $V_{0}$.
There also exists an infinite set of equivalence classes
$\mathcal{E}_{\psi}$ themselves.  If $|\psi_{1} \rangle = (\alpha_{1},
\beta_{1} )^{T}$ and $|\psi_{2} \rangle = (\alpha_{2}, \beta_{2} )^{T}$
are such that $U_{I} |\psi_{1} \rangle \neq |\psi_{2} \rangle$ for any
choice of $V_{0}$, then the two equivalence classes
$\mathcal{E}_{\psi_{1}}$ and $\mathcal{E}_{\psi_{2}}$ are distinct and
have no common states. Theoretically there is no preference as to what
equivalence class to use for MLE of a qubit, however, in a practical
situation some specific equivalence class will be selected, for example,
depending on the initial state of the qubit physical system or the
technical limitations of the available controls.

\subsection{The Pauli matrices}

We will denote the high-dimensional representations of the Pauli
matrices $\{ \sigma_{x}, \sigma_{y}, \sigma_{z} \}$ as $\{ U_{x}, U_{y},
U_{z} \}$. The construction of these operations for a multilevel-encoded
qubit is similar to that of $U_{I}$.  In particular, the form of the
bit-flip operation $U_{x}$ (the NOT gate) under the condition of logical
equivalence is
\begin{equation}
\label{Ux}
U_{x} = \left( \begin{array}{cc} 0 & V_{1} \\ V_{1} & 0
\end{array} \right)
= \sigma_{x} \otimes V_{1},
\end{equation}
where $V_{1}$ is an arbitrary $n \times n$ unitary matrix.  It is easy
to verify that
\begin{equation}
\label{Ux-Eq}
U_{x} |\psi\rangle = \left( \begin{array}{c} V_{1} \beta \\ V_{1} \alpha
\end{array} \right) \sim
\left( \begin{array}{c} \beta \\ \alpha
\end{array} \right).
\end{equation}
As demonstrated in (\ref{Ux-Eq}), $U_{x}$ performs the desired bit-flip
operation.  Analogously, the other two Pauli matrices $U_{y}$ and
$U_{z}$ (the latter being the phase-flip operation) have the form
\begin{equation}
U_{y} = \left( \begin{array}{cc} 0 & - i V_{2} \\ i V_{2} & 0
\end{array} \right)
= \sigma_{y} \otimes V_{2}
\end{equation}
and
\begin{equation}
\label{Uz}
U_{z} = \left( \begin{array}{cc} V_{3} & 0 \\ 0 & -V_{3} \end{array}
\right) = \sigma_{z} \otimes V_{3},
\end{equation}
where $V_{2}$ and $V_{3}$ are also arbitrary $n \times n$ unitary
matrices. Under logical equivalence, the multiplication rule $\sigma_{j}
= i \epsilon_{j k l} \sigma_{k} \sigma_{l}$ (where $\{ j,k,l \} = \{
x,y,z \}$ and $\epsilon_{j k l}$ is the completely antisymmetric unit
tensor) and the anti-commutation rule $\{\sigma_{j}, \sigma_{k} \} = 0$
$(j \neq k)$ for the Pauli matrices are expressed as
\begin{subequations}
\begin{gather}
U_{j} \sim i\epsilon_{j k l} U_{k} U_{l}, \ \  \{ j,k,l \} = \{  x,y,z
\} \\ U_{j} U_{k} \sim - U_{k} U_{j} \ \ (j \neq k),
\end{gather}
\end{subequations}
respectively.  One can easily verify that these relations are satisfied
given the form of $U_{x}$, $U_{y}$ and $U_{z}$ and the arbitrary nature
of $V_{1}$, $V_{2}$, and $V_{3}$.

\subsection{The Hadamard gate}

We denote the high-dimensional representation of the Hadamard gate $H$
as $U_{H}$.  Since $H \sigma_{z} H = \sigma_{x}$, $U_{H}$ must satisfy
$U_{H} U_{z} U_{H} |\psi\rangle \sim U_{x}|\psi\rangle$ for any
$|\psi\rangle$. Consider $U_{H}$ of the most general form of
(\ref{U-mas}):
\begin{equation}
\left( \begin{array}{ccc} A V_{3} A - B V_{3} C\, & A V_{3} B - B V_{3}
D \\ C V_{3} A - D V_{3} C\, &  C V_{3} B - D V_{3} D
\end{array} \right) \!\!
\left( \begin{array}{c} \alpha \\ \beta
\end{array} \right) \sim
\left( \begin{array}{c} \beta \\ \alpha
\end{array} \right),
\end{equation}
which is true if $A = B = C = - D$ and $\sqrt{2} A$ is unitary.
Therefore, $U_{H}$ has the form
\begin{equation}
\label{H-gate}
U_{H} = \frac{1}{\sqrt{2}} \left( \begin{array}{cr} V_{4} & V_{4} \\
V_{4} & -V_{4}
\end{array} \right)
= H \otimes V_{4},
\end{equation}
where $V_{4}$ is an arbitrary $n \times n$ unitary matrix.

\subsection{The phase-shift gate}

We denote the high-dimensional representation of the phase-shift gate
$P(\phi)$ as $U_{P} (\phi)$, which has the form
\begin{equation}
\label{P-gate}
U_{P} (\phi) = \left( \begin{array}{cc} V_{5} & 0 \\ 0 & V_{5} e^{i \phi}
\end{array} \right)
= P(\phi) \otimes V_{5},
\end{equation}
where $\phi \in \mathbb{R}$ and $V_{5}$ is an arbitrary $n \times n$
unitary matrix.

\subsection{Weak commutation}

Under logical equivalence, we can introduce a \emph{weak} commutation
relation.  Two operators $F$ and $G$ are said to weakly commute if
\begin{equation}
F G \sim G F.
\end{equation}
Within the MLE formalism, every logical operation should weakly commute
with $U_{I}$ (the weak identity).  It is not difficult to see that
$U_{I}$ weakly commutes with the logical operations $U_{x}$, $U_{y}$,
$U_{z}$, $U_{H}$ and $U_{P} (\phi)$ described above. Moreover, we show
that $U_{I}$ weakly commutes with any unitary operation $U_{\xi}$ such
that
\begin{equation}
U_{\xi} \in \text{U}(2) \otimes \text{U}(n),
\end{equation}
which is a subgroup of U($2n$).  In order to prove that
\begin{equation}
\label{w-comm}
U_{I} U_{\xi} \sim U_{\xi} U_{I},
\end{equation}
first note that any such operation can be written as $U_{\xi} = \xi
\otimes V$, where $\xi \in \text{U}(2)$ and $V \in \text{U}(n)$.  Then
we find that
\begin{equation}
\label{proof1}
\langle\psi| U_{\xi}^{\dagger} U_{I}^{\dagger} \Lambda_{r} U_{I}
U_{\xi}|\psi\rangle = \text{Tr} \left( \left( \xi^{\dagger}\lambda_{r}
\xi \right) \otimes \openone_{n} |\psi\rangle\langle\psi| \right)
\end{equation}
and
\begin{equation}
\label{proof2}
\langle\psi| U_{I}^{\dagger} U_{\xi}^{\dagger} \Lambda_{r} U_{\xi} U_{I}
|\psi\rangle = \text{Tr} \left( \left( \xi^{\dagger}\lambda_{r} \xi
\right) \otimes \openone_{n} |\psi\rangle\langle\psi| \right),
\end{equation}
where
\begin{equation}
V_{0}^{\dagger} V^{\dagger} V V_{0} = V^{\dagger} V_{0}^{\dagger} V_{0}
V = \openone_{n}
\end{equation}
for any unitary $V_{0}$ and $V$.  Since the right-hand sides of
(\ref{proof1}) and (\ref{proof2}) are equal, their left-hand sides must
be equal as well, which concludes the proof of (\ref{w-comm}).
Additionally, if an operation $U_{\xi}$ weakly commutes with $U_{I}$,
i.e., $U_{I} U_{\xi} \sim U_{\xi} U_{I}$, then $U_{\xi}$ is an element
of $\text{U}(2) \otimes \text{U}(n)$.

\subsection{Comments and remarks}

Due to the arbitrary nature of the encoded component $V$ ($V \in
\text{U}(n)$), every logical operation is represented in the
$(2n)$-dimensional space by an infinite number of logically equivalent
transformations $U_{\xi} = \xi \otimes V$ (which collectively form an
equivalence class characterized by $\xi$).  Therefore, the use of MLE of
logical basis states allows for a great flexibility in the choice of the
actual physical transformation that realizes the desired logical
operation, which can be an important advantage in many practical
situations.

An interesting question is whether the flexibility in the encoded
component of $U_{\xi}$ makes it easier to find an optimal control field
that produces a physical transformation belonging to the desired
equivalence class of logical operations.  The answer is not yet clear
since $\xi$, the logical component of $U_{\xi}$, must still  be produced
with the same accuracy as in the single-level representation.  Another
question for future research is whether the flexibility due to the
logical equivalence of different physical transformations improves the
robustness of logical operations against noise in controls.

A restriction in the MLE formalism presented above is the requirement
that both logical basis states, $|0\rangle$ and $|1\rangle$, are encoded
by the same number of physical levels.  Satisfying this requirement in a
real physical system can sometimes be a difficult task.  Consider for
example, a logical basis encoded into vibrational levels on two
electronic surfaces of a molecule.  If the difference between the two
vibrational frequencies is sufficiently large, a laser pulse of a given
spectral width will encompass a different number of vibrational levels
on the two surfaces.  Therefore, it is necessary to consider how
restricting the encoding subspaces to be of equal dimension can be dealt
with in practice.  First note that in any practical situation it is
impossible to achieve the target quantum operation \emph{exactly}.
Correspondingly, laboratory controls are designed to produce an
operation which is as close as possible to the target (an optimal
control problem).  In the case of SLE, the actual quantum operation
should be as close as possible to the $2 \times 2$ target operation
$\xi$.  Analogously, in the case of MLE, the actual quantum operation
should be as close as possible to the $(2n) \times (2n)$ target
operation $U_{\xi} = \xi \otimes V$ (with the additional flexibility
provided by the arbitrary nature of $V$).  Consider now what will happen
if in reality one logical basis state is encoded into $n$ levels and the
other into $m = n+k$ levels ($k > 0$).  In such a case the actual
operation will be represented by an $(n+m) \times (n+m)$ unitary matrix
$U_{\text{lab}}$, and the target operation should be of the form
\begin{equation}
\label{u-target-nm}
U_{\text{target}} = (\xi \otimes V) \oplus W,
\end{equation}
where $W$ is an arbitrary $k \times k$ unitary matrix (this mathematical
structure also frequently appears in the context of QEC and operator
QEC, e.g., see \cite{KnLa97,KLP05}).  The physical meaning of this
target operation is that a qubit encoded in the $(2n)$-dimensional space
should not couple to the additional $k$ levels.  So the laboratory task
is to design controls which will produce the actual operation
$U_{\text{lab}}$ as close as possible to $U_{\text{target}}$ of
(\ref{u-target-nm}).  This is an optimal control problem which is not
different \emph{in principle} from the one encountered when $n = m$.  Of
course, the practical difficulty of finding optimal control fields can
increase when the encoding subspaces have different dimensions, but the
method of closed-loop laboratory control with shaped laser pulses is
well-suited for dealing with such problems.  Moreover, one can argue
that optimal control could benefit from freedom in the choice of the
actual parceling of the vibrational levels between the two subspaces
since the experimenter does not have to \emph{a priori} define the
levels into pre-assigned groups.  The most effective partitioning of
physical levels into the logical basis can be included as part of the
optimal control problem, with this additional freedom providing a
potential for better solutions.

\section{Multilevel Encoding for Mixed States}
\label{sec:dense}

The principle of logical equivalence can be extended to mixed states.
Consider a density matrix that represents a state in the expanded $(2
n)$-dimensional Hilbert space:
\begin{equation}
\label{density}
\rho = \sum_{i,j = 1}^{2} \left( r_{i j}|i\rangle_\mathrm{L}\,
{}_\mathrm{L}\!\langle j| \right) \otimes R_{i j} = \left(
\begin{array}{cc} r_{11} R_{11} & r_{12} R_{12} \\ r_{21} R_{21} &
r_{22} R_{22}
\end{array} \right),
\end{equation}
where $r_{i j}$ are matrix elements representing the state of the
logical component and $R_{i j}$ are $n \times n$ matrices representing
the state of the encoded component.  As a density matrix, $\rho$ has
certain fundamental properties: $\rho = \rho^{\dagger}, ~
\langle\psi|\rho|\psi\rangle \geq 0$ for all $|\psi\rangle$, and
$\text{Tr} \left( \rho \right) = 1$.  In \ref{app:a} we show that the
logical-component matrix
\begin{equation}
\rho_{\mathrm{L}} = \left( \begin{array}{cc} r_{11} & r_{12} \\ r_{21} &
r_{22}
\end{array} \right)
\end{equation}
as well as the diagonal sub-blocks $R_{11}$ and $R_{22}$ also have these
properties and thus are proper density matrices.

Logical equivalence for two density matrices $\rho$ and $\rho'$ is
defined as
\begin{equation}
\label{eq:le-rho} \text{Tr} \left( \rho \Lambda_{r} \right) =
\text{Tr} \left( \rho' \Lambda_{r} \right), \ \ r = \{x,y,z\},
\end{equation}
plus the normalization condition $\text{Tr} \left( \rho \right) =
\text{Tr} \left( \rho' \right) = 1$.  Two unitary operations $U_{1}$ and
$U_{2}$ are logically equivalent if they satisfy
\begin{equation}
U_{1} \rho U_{1}^{\dagger} \sim U_{2} \rho U_{2}^{\dagger}.
\end{equation}
By evaluating the corresponding traces, it is straightforward to show
that for the weak identity $U_{I}$,
\begin{equation}
\rho \sim U_{I} \rho U_{I}^{\dagger}
\end{equation}
and
\begin{gather}
\nonumber U_{I} U_{\xi} \rho U_{\xi}^{\dagger} U_{I}^{\dagger} \sim
U_{\xi} U_{I} \rho U_{I}^{\dagger} U_{\xi}^{\dagger} \\ \text{for any} \
U_{\xi} \in \text{U}(2) \otimes \text{U}(n),
\end{gather}
which is completely analogous to the result obtained for pure states in
the previous section.

Now we investigate the process of \emph{non-unitary} evolution (e.g.,
due to environmentally-induced decoherence) in the context of MLE.  We
restrict our consideration to processes which are unitary in the logical
component and general (not necessarily unitary) in the encoded
component.  Quantum operations (acting on the reduced density matrix
$\rho$ of the system of interest) representing such processes will be
symbolically denoted as $\mathcal{L}_{\xi} \equiv \{ \xi, \mathcal{W}
\}$, where $\xi$, the logical component, is a $2 \times 2$ unitary
matrix and $\mathcal{W}$, the encoded component, is a generalized
quantum operation whose properties will be determined below.  The action
of $\mathcal{L}_{\xi}$ on $\rho$ is represented as
\begin{equation}
\label{eq:L} \mathcal{L}_{\xi} \rho = \sum_{i,j = 1}^{2} \xi
\left( r_{i j}|i\rangle_\mathrm{L}\, {}_\mathrm{L}\!\langle j| \right)
\xi^{\dagger} \otimes \left( \mathcal{W} R_{i j} \right).
\end{equation}
Using the explicit matrix form of $\xi$, this can be rewritten as
\begin{equation}
\mathcal{L}_{\xi} \rho = \sum_{i,j,k,l = 1}^{2} \left( \xi_{i j} r_{j k}
\xi_{l k}^{*} |i\rangle_\mathrm{L}\, {}_\mathrm{L}\!\langle l| \right)
\otimes \left( \mathcal{W} R_{j k} \right).
\end{equation}
Since $\mathcal{L}_{\xi}$ and $\mathcal{W}$ act on the density matrices
$\rho$ and $ \{ R_{ii} \}$, respectively, these operations must preserve
Hermiticity, positivity, and the trace.  Therefore, a permissible
operation $\mathcal{W}$ is a positive and trace-preserving map.

Next, consider the effect of the generalized weak identity,
$\mathcal{L}_{I} = \{\openone_{2}, \mathcal{W}\}$, on $\rho$:
\begin{equation}
\label{eq:LI} \rho' = \mathcal{L}_{I} \rho = \sum_{i,j = 1}^{2}
\left( r_{i j}|i\rangle_\mathrm{L}\, {}_\mathrm{L}\!\langle j| \right)
\otimes \left( \mathcal{W} R_{i j} \right).
\end{equation}
Using the explicit form of the EA operators,
\begin{equation}
\Lambda_{r} = \lambda_{r} \otimes \openone_{n} = \sum_{i,j = 1}^{2}
\lambda^{(r)}_{i j} |i\rangle_\mathrm{L}\, {}_\mathrm{L}\!\langle j|
\otimes \openone_{n},
\end{equation}
we obtain
\begin{align}
\nonumber \text{Tr} \left( \rho' \Lambda_{r} \right) & = \sum_{i,j =
1}^{2} r_{i j} \lambda^{(r)}_{j i} \text{Tr} \left( \mathcal{W} R_{i j}
\right) \\ & = \sum_{i,j = 1}^{2} r_{i j} \lambda^{(r)}_{j i} \text{Tr}
\left( R_{i j} \right) = \text{Tr} \left( \rho \Lambda_{r} \right),
\end{align}
where we used the fact that $\mathcal{W}$ is a trace-preserving
operation.  This proves that $\mathcal{L}_{I}$ indeed has the property
of the weak identity:
\begin{equation}
\label{LE:L}
\mathcal{L}_{I} \rho \sim \rho.
\end{equation}
An immediate and interesting consequence of (\ref{LE:L}) is the logical
equivalence of mixed and pure states.  In other words, while
$\rho_{\psi} = |\psi\rangle\langle\psi|$ represents a pure state in the
form of (\ref{qubit}), its logically equivalent state $\rho' =
\mathcal{L}_{I} \rho_{\psi}$ will be mixed if $\mathcal{L}_{I}$ includes
any permissible and non-unitary operation $\mathcal{W}$.  Thus, a
logical basis state can be encoded into a subspace of multiple physical
levels which are mixed (e.g., by a thermal excitation) without any loss
of coherence \emph{between} the two logical basis states that form a
qubit. For example, if an initial state is such that mixing is only
\emph{within} an encoding subspace and not \emph{between} the two
subspaces, then quantum computation with MLE can be conducted without
any loss of fidelity due to intra-subspace mixing.  This property makes
the method of MLE a very promising practical alternative to the
cumbersome process of ground-state cooling.

An additional property of $\mathcal{W}$ is found by examining the weak
commutation relation.  Applying the method of proof used in
(\ref{proof1}) and (\ref{proof2}) to general quantum operations in the
form of (\ref{eq:L}), it follows that any permissible operation
$\mathcal{L}_{\xi}$ weakly commutes with $\mathcal{L}_{I}$, i.e.,
\begin{equation}
\label{w-comm:L}
\mathcal{L}_{I} \mathcal{L}_{\xi} \rho \sim \mathcal{L}_{\xi}
\mathcal{L}_{I} \rho,
\end{equation}
if
\begin{equation}
\label{W-closure}
\text{Tr} \left( \mathcal{W}_{0} \mathcal{W} R_{i j} \right) = \text{Tr}
\left( \mathcal{W} \mathcal{W}_{0} R_{i j} \right) \ \ \text{for all} \
i, \, j.
\end{equation}
Here, $\mathcal{W}_{0}$ and $\mathcal{W}$ are the encoded components of
$\mathcal{L}_{I}$ and $\mathcal{L}_{\xi}$, respectively.  Equations
(\ref{w-comm:L}) and (\ref{W-closure}) show that all permissible
operations $\{ \mathcal{W} \}$ form a closed set.  In other words, if
$\mathcal{W}$ and $\mathcal{W}'$ are two arbitrary permissible
operations, then $\mathcal{W}'' = \mathcal{W} \mathcal{W}'$ is another
permissible operation.  An example of a positive and trace-preserving
map with this closure property is the Kraus representation:
\begin{equation}
\mathcal{W} R_{i j} = \sum_{\nu} W_{\nu} R_{i j} W_{\nu}^{\dagger} , \ \
\text{where} \ \sum_{\nu} W_{\nu}^{\dagger} W_{\nu} = \openone_{n} ,
\end{equation}
which is in fact a \emph{completely} positive map \cite{NiCh00}.

It is now possible to generalize the notion of a class of logically
equivalent operations to non-unitary processes.  For each logical
component $\xi \in \text{U}(2)$, there exists an infinite number of
equivalent logical operations $\mathcal{L}_{\xi}$ (forming an
equivalence class) which permit decoherence \emph{within} an encoding
subspace but preserve coherence \emph{between} the encoding subspaces.
Specifically, any two operations $\mathcal{L}_{\xi} = \{ \xi,
\mathcal{W} \}$ and $\mathcal{L'}_{\xi} = \{ \xi, \mathcal{W}' \}$, with
permissible maps $\mathcal{W}$ and $\mathcal{W'}$, are logically
equivalent (also illustrating that a non-unitary operation
$\mathcal{L}_{\xi} = \{ \xi, \mathcal{W} \}$ is logically equivalent to
a unitary operation $U_{\xi} = \xi \otimes V$, where $V \in
\text{U}(n)$).  Accordingly, MLE allows for quantum computation with
full fidelity in the presence of non-unitary evolution within the
encoding subspaces.

\section{Example: Numerical simulation of closed-loop optimal control
  of unitary operations in a multilevel system}
\label{sec:example}

To illustrate the formalism and advantages of MLE, as developed in
sections~\ref{sec:concept}-\ref{sec:dense}, we perform a simulation of
closed-loop optimal control of single-qubit unitary operations for a
model system.  It should be emphasized that the method of closed-loop
optimal control is uniquely suited for laboratory implementation, but is
very difficult to simulate on the computer, except for the simplest
models.  This is related to the fact that learning algorithms (e.g.,
genetic algorithms) which are used to search for the optimal control
fields, involve vast amounts of data processing.  In our numerical
simulations, solving the time-dependent Schr\"{o}dinger equation for
each set of control parameters is the limiting computational step. Since
a learning algorithm typically needs to search over hundreds of
thousands or even millions of points in the parameter space until a
global optimum is found, the simulation can take days even for
relatively simple model systems. However, in the laboratory, a real
physical system, whatever complex it may be, ``solves'' its own
time-dependent Schr\"{o}dinger equation in real time.  Hence, the data
processing of the learning algorithm in the laboratory is limited mainly
by the repetition rate of the laser and pulse-shaping system, which can
be as fast as several kilohertz and consequently, the search over a
million of different control fields can be completed in just several
minutes.

\subsection{The model}

Due to the previous considerations, we perform just a proof-of-principle
numerical simulation of optimal control for a simple four-level model
system which nevertheless demonstrates the difference between the MLE
and SLE operations in the presence of decoherence. This model represents
a single qubit with each logical basis state encoded either by two
levels in the case of MLE ($n = 2$) or by one level in the case of SLE
(the other two levels remain unused). The energy spacings correspond to
the electronic and vibrational levels of the sodium dimer,
$\text{Na}_{2}$.\footnote{The energy separation between the ground and
first excited electronic surfaces is 0.066889653 a.u.; the separations
between vibrational levels are 0.0007250238 a.u. and 0.000534563 a.u. on
the ground and excited electronic surfaces, respectively.  The data are
taken from \cite{H-numbers}.} For MLE, the ground and excited electronic
states correspond to the $|0\rangle$ and $|1\rangle$ logical basis
states, respectively, i.e., each logical basis state is encoded by a
pair of vibrational levels on the corresponding electronic surface.  For
SLE, the logical basis states $|0\rangle$ and $|1\rangle$ are defined as
the lower vibrational levels on the ground and excited electronic
surfaces, respectively.  In a homonuclear dimer such as $\text{Na}_{2}$,
transitions between vibrational levels within a given electronic surface
are forbidden (the dipole moment for these transitions is zero).
Transitions between different electronic surfaces are allowed and
correspond to the dipole moment set to unity.  Thus, the number of
allowed transitions equals $n^{2}$.

Unary multilevel atomic and molecular systems, such as $\text{Na}_{2}$,
are not scalable for quantum computing applications in the absence of
physical entanglement.  Implementations of quantum algorithms without
entanglement have been investigated using Rydberg atoms \cite{Rydberg-a,
Rydberg-b} and linear optical systems \cite{NJC98, PGK00, LP04}.  It was
concluded that quantum computation with a single multilevel system is
possible, but requires exponentially greater overhead than a
multiparticle quantum computer that operates with entanglement
\cite{SL00, DAM00}.  Therefore, there exists significant interest in
engineering interactions between atoms or molecules for creating
entanglement needed for efficient and scalable quantum information
processing \cite{GKB99, GKB00, GKB02, DJ99, TC00, TC05, DD02, DV05,
MS05, LC05}.  However, in our example scalability is not the issue, as
we use the model multilevel system to investigate some of the most basic
features of MLE for a single qubit.

There exist plans to use homonuclear dimers such as $\text{Na}_{2}$ and
$\text{K}_{2}$ for experimentally studying the basic functioning of MLE
in single-qubit systems in the presence of thermalization.  These
experiments would benefit from existing technologies for managing and
measuring the dynamics of vibrational molecular wavepackets
\cite{WaWa98, TJD93}, including applications of closed-loop control
methods \cite{MB06}.  A much greater challenge would be an experimental
realization of MLE in entangled multi-qubit systems, such as photon-atom
quantum memories, as the technology for this future type of experiments
is not yet fully developed.

\subsection{Formulation of the optimal control problem}

The time evolution operator $U(t)$ for an isolated quantum system
satisfies the Schr\"{o}dinger equation:
\begin{equation}
\label{Schro}
\frac{d}{d t} U(t) = -\frac{i}{\hbar} H(t) U(t).
\end{equation}
The time-dependent Hamiltonian, $H(t)$, in this model is
\begin{equation}
\label{Hamilton}
H(t) = H_{0} - \mu\epsilon(t),
\end{equation}
where $H_{0}$ is the system Hamiltonian in the absence of control, $\mu$
is the electric dipole-moment operator and $\epsilon(t)$ is the
time-dependent control field, defined as
\begin{equation}
\label{efield}
\epsilon(t) = f(t)\sum_{i = 1}^{n^{2}} a_{i} \cos(\omega_{i}t +
\delta_{i}).
\end{equation}
Here, $f(t)$ is an envelope function which incorporates the laser pulse
width (e.g., a Gaussian or similar type of distribution) and $a_{i}$,
$\omega_{i}$, $\delta_{i}$ are the amplitude, frequency, and relative
phase of the $i$th electric field component, respectively. Transition
frequencies are determined by the system Hamiltonian.  The operation is
over the time interval $[0, t_{f}]$.\footnote{In our simulations,
$t_{f}$ can vary from $2 \times 10^{3}$ to $5 \times 10^{4}$ a.u. (i.e.,
from about 48 fs to 1.2 ps).}

The control objective is to achieve the time evolution operator $U(
t_{f} )$ which is as close as possible to the target transformation
$U_{\text{target}}$.  The fitness of the control field is evaluated by
using the gate fidelity, which is a functional of control:
\begin{equation}
\label{norm}
\mathcal{F}[\epsilon(t)] = 1 - \left\| U_{\text{target}} - U(t_{f})
\right\| .
\end{equation}
Optimal control solutions correspond to maxima of $\mathcal{F}$.  The
search for a global maximum is performed using a genetic algorithm (GA)
implemented with population sizes of $\sim 200$, several different
reproductive schemes, and crossover and mutation rates between 20 and 40
percent.

\subsubsection{Fidelity functionals for MLE.}

For MLE, the specific matrix norm in (\ref{norm}) will depend on the
target transformation.  The general form of the unitary evolution
operator $U(t_{f})$ can be written as
\begin{equation}
U(t_{f}) = \left( \begin{array}{cc} A & B\\ C & D
\end{array} \right),
\end{equation}
where $A, B, C,$ and $D$ are square matrices of dimension $n$. For the
bit-flip operation $U_{x}$ of (\ref{Ux}) and the phase-flip operation
$U_{z}$ of (\ref{Uz}) as the target transformations, we use fidelity
functionals of the form
\begin{equation}
\label{norm-x}
\mathcal{F}_{x} = 1 - \left( \frac{ \left\| BB^{\dagger} - 1
\right\|^{2}}{n} + \frac{ \left\| B - C \right\|^{2}}{4n} \right)^{1/2}
\end{equation}
and
\begin{equation}
\label{norm-z}
\mathcal{F}_{z} = 1 - \left(\frac{ \left\| AA^{\dagger} - 1
\right\|^{2}}{n} + \frac{ \left\| A + D \right\|^{2}}{4n} \right)^{1/2} ,
\end{equation}
respectively, where the matrix norm is defined as $\left\| M
\right\|^{2} = \text{Tr} \left( M M^{\dagger} \right)$ and the
coefficients are chosen for the proper normalization, so that $0 \leq
\mathcal{F}_{x} \leq 1$ and $0 \leq \mathcal{F}_{z} \leq 1$.  The
maximum of the fidelity functional ($\mathcal{F}_{x} = 1$ or
$\mathcal{F}_{z} = 1$) is achieved when the actual transformation
$U(t_{f})$ is an element of the corresponding equivalence class ($U_{x}$
or $U_{z}$).

\subsubsection{Fidelity functionals for SLE.}

For SLE, two vibrational levels are used to encode the logical basis
states, while the other two levels are unused.  Therefore, the bit-flip
and the phase-flip operations are represented by the unitary
transformations
\begin{equation}
\label{Ux-SLE} U'_{x} = \sigma_{x} \oplus v
\end{equation}
and
\begin{equation}
\label{Uz-SLE} U'_{z} = \sigma_{z} \oplus w ,
\end{equation}
respectively, where $\sigma_{x}$ and $\sigma_{z}$ are the corresponding
Pauli matrices acting on the space of the logical basis, and $v$ and $w$
are arbitrary $2 \times 2$ unitary matrices acting on the space of the
two unused levels.  For computations, we rearrange $U'_{x}$ and $U'_{z}$
so that their indices correspond to the physical order of the energy
levels (the logical basis states $|0\rangle$ and $|1\rangle$ correspond
to the 1st and 3rd energy levels, respectively).  For the SLE operations
$U'_{x}$ and $U'_{z}$ as the target transformations, the fidelity
functional will be of the form
\begin{equation}
\label{norm-x-SLE} \mathcal{F}'_{x} = \frac{1}{2} \left| U_{13} +
U_{31} \right|
\end{equation}
and
\begin{equation}
\label{norm-z-SLE} \mathcal{F}'_{z} = \frac{1}{2} \left| U_{11} -
U_{33} \right| ,
\end{equation}
respectively, where $U_{i j}$ denotes the corresponding matrix element
of the actual transformation $U( t_{f} )$.  The maximum of the fidelity
functional ($\mathcal{F}'_{x} = 1$ or $\mathcal{F}'_{z} = 1$) is
achieved when the actual transformation $U(t_{f})$ produces the target
transformation ($\sigma_{x}$ or $\sigma_{z}$, up to a global phase) in
the two-level subspace of the logical basis.

\subsection{Numerical optimization and analysis of results}

Table \ref{fitness} summarizes the gate fidelities obtained from the GA
optimization for MLE ($n=2$) and SLE configurations of the four-level
model system.  The GA is capable of finding optimal control solutions
for both MLE and SLE cases with reasonable accuracies, limited mainly by
the parameterized structure of the control fields in (\ref{efield}).
These fidelities could be improved by coupling the results of the GA to
an additional gradient-based search algorithm that does not rely on any
particular parameterization of the control fields. However, for the
purposes of our simulations, the logical operations obtained with the GA
were sufficient to demonstrate the main distinctions between MLE and SLE.

We are also interested to see what happens when we apply the optimal
transformation located by GA optimization on an initial state which is
mixed by a decoherence process (we assume that the effect of decoherence
\emph{during} the logical operation is negligible and that only the
initial state is affected).  We consider two types of decoherence
processes: dephasing (only for MLE) and thermalization (for both MLE and
SLE).
\begin{table}
\caption{Fidelities of MLE and SLE logical operations using the GA
optimization.} \label{fitness}
\begin{indented}
\item[]\begin{tabular}{@{}cccc} \br $\mathcal{F}_{x}$ &
$\mathcal{F}_{z}$ & $\mathcal{F}'_{x}$ & $\mathcal{F}'_{z}$ \\ \mr \
0.9987 \ & \ 0.9975 \ & \ 0.9998 \ & \ 0.9996 \  \\ \br
\end{tabular}
\end{indented}
\end{table}

To investigate the effect of initial state decoherence on the fidelity
of quantum operations, we consider the difference between the perfect
(or intended) initial state rotated by the perfect target transformation:
\begin{equation}
\rho_{\text{target}} = U_{\text{target}} \rho
U_{\text{target}}^{\dagger} ,
\end{equation}
and the actual initial state rotated by the optimal actual
transformation found by the GA:
\begin{equation}
\rho(t_{f}) = U(t_{f}) \tilde{\rho} U(t_{f})^{\dagger} .
\end{equation}
The actual initial state $\tilde{\rho}$ may be affected by a decoherence
process and therefore  differ from the perfect initial state $\rho$.
For MLE, the difference between $\rho_{\text{target}}$ and $\rho(t_{f})$
is estimated by evaluating the ``error of equivalence'':
\begin{equation}
\label{error-MLE}
\varepsilon [\rho,\tilde{\rho}] = \frac{1}{3} \sum_{r = x, y, z} \left|
\text{Tr} \left( \Lambda_{r} \left[ \rho(t_{f}) - \rho_{\text{target}}
\right] \right) \right| ,
\end{equation}
where $\{ \Lambda_{x} , \Lambda_{y} , \Lambda_{z} \}$ are the three EA
operators of (\ref{EA-ops}).  This error measures how far the two
density matrices are from complete logical equivalence.  The worst value
of $\varepsilon$ is 1, while in the case of complete equivalence
$\varepsilon = 0$.  For SLE, the distance between the perfect target
state and the actual final state is measured by
\begin{equation}
\label{error-SLE}
\varepsilon' [\rho,\tilde{\rho}] = \text{Tr} \left( \left[ \rho(t_{f}) -
\rho_{\text{target}} \right]^{2} \right) .
\end{equation}
In the absence of decoherence ($\tilde{\rho} = \rho$), the errors
$\varepsilon$ and $\varepsilon'$ will be non-zero only due to the fact
that the optimal control solution found by the GA produces a
transformation that is slightly different from the target one.  However,
if decoherence impairs the initial state ($\tilde{\rho} \neq \rho$),
this can significantly affect the error.  We will see below that
operations resulting from MLE are not at all hindered by dephasing or
thermalization of the initial state.  In contrast, the performance of
SLE operations deteriorates if the initial state is thermalized.

\subsubsection{MLE operations in the presence of dephasing.}

Dephasing of non-stationary vibrational wave packets in hot alkaline
dimers is caused by the vibrational-rotational coupling
\cite{BRWW01}. The typical dephasing time is inversely proportional to
the temperature and varies from $\sim 3$ ps for very hot molecules
(produced in a heat pipe at about 400 ${}^{\circ}$C) to $\sim 20$ ps for
vapours cooled to about 100~K.  Note that the decoherence time on the
electronic surfaces (i.e., \emph{within} the encoding subspaces of the
MLE scheme) due to this coupling is much shorter than that
\emph{between} electronic surfaces associated with spontaneous emission
from the excited electronic surface, which happens on the time-scale of
several nanoseconds. Therefore, dephasing induced by the
vibrational-rotational coupling is the most important mechanism of
decoherence for vibrational wave-packets.  Since the length of the
optimal control pulse does not exceed 1.2 ps (and can even be made
shorter at the cost of a small decrease in the gate fidelity), the
effect of dephasing \emph{during} the logical operation is negligible at
temperatures about 100~K.

We study the effect of dephasing on MLE operations by considering a set
of randomly generated pure initial states of the form
\begin{equation}
\label{psi-init}
|\psi \rangle = c_{0} |0\rangle_{\text{L}} \otimes \left(
\begin{array}{c} \cos \left( \theta_{0} \right) \\ \exp \left( i\phi_{0}
\right)\sin \left( \theta_{0} \right)
\end{array} \right)
+ c_{1} |1\rangle_{\text{L}} \otimes \left( \begin{array}{c} \cos \left(
\theta_{1} \right) \\ \exp \left( i\phi_{1} \right)\sin \left(
\theta_{1} \right)
\end{array} \right),
\end{equation}
where $c_{0}, \, c_{1}, \, \phi_{0}, \, \phi_{1}, \, \theta_{0}$, and
$\theta_{1}$ are randomly generated real parameters such that
\begin{subequations}
\begin{gather}
c_{0}^{2} + c_{1}^{2} = 1 , \\ 0 \, \leq \, \phi_{0}, \, \phi_{1}, \,
\theta_{0}, \, \theta_{1} \, \leq \, 2\pi .
\end{gather}
\end{subequations}
The initial state $|\psi \rangle$ corresponds to the density matrix
$\rho = | \psi \rangle\langle \psi |$, whose elements in the basis of
the four energy levels are denoted as $\langle i | \rho | j \rangle =
r_{i j}$ ($i,j = 1,2,3,4$).  Dephasing within each encoding subspace is
represented by setting the off-diagonal matrix elements of each diagonal
sub-block equal to zero, resulting in the mixed state
\begin{equation}
\tilde{\rho} = \left( \begin{array}{cccc} r_{11} & 0 & r_{13} & r_{14}
\\ 0 & r_{22} & r_{23} & r_{24} \\ r_{31} & r_{32} & r_{33} & 0 \\
r_{41} & r_{42} & 0 & r_{44} \\
\end{array} \right).
\end{equation}
The error of equivalence for the $U_{x}$ and $U_{z}$ target
transformations is evaluated both in the absence and in the presence of
dephasing: $\varepsilon [\rho,\rho]$ and  $\varepsilon
[\rho,\tilde{\rho}]$, as defined in (\ref{error-MLE}), and averaged over
$10^{6}$ random choices of the initial state $|\psi\rangle$ in the form
of (\ref{psi-init}).  These average errors, $\langle \varepsilon
\rangle$, are presented in table~\ref{error-dephase}.  We see that the
operation error does not increase (and even slightly decreases) when the
initial state is affected by dephasing.  This means that the pure state
$\rho$ and the dephased state $\tilde{\rho}$ are equally suitable for
MLE operations due to their logical equivalence.

\begin{table}
\caption{Average errors of MLE operations with and
without dephasing of the initial state.}
\label{error-dephase}
\begin{indented}
\item[]\begin{tabular}{@{}ccc} \br \ Initial state \ & \ $\langle
\varepsilon \rangle \ \text{for} \ U_{x}$ \ & \ $\langle \varepsilon
\rangle \ \text{for} \ U_{z}$ \ \\ \mr $\rho$ & \ 0.0093 \  & \ 0.0146 \
\\ $\tilde{\rho}$ & \ 0.0076 \  & \ 0.0134 \ \\ \br
\end{tabular}
\end{indented}
\end{table}

\subsubsection{MLE and SLE operations in the presence of
thermalization.}

As mentioned above, alkaline dimers are typically produced at high
temperatures of up to 400 ${}^{\circ}$C.  Moreover, these molecules are
unstable at low temperatures, although vapours of sufficiently low
concentrations can be carefully cooled to 100~K or even slightly
below. We study the effect of thermalization on both MLE and SLE
operations by considering the following pure initial state:
\begin{equation}
\label{density2} |\psi \rangle = |0\rangle_{\text{L}} \otimes
\left( \begin{array}{c} 1 \\ 0 \end{array} \right) = (1, 0, 0, 0)^{T} ,
\end{equation}
which corresponds to the lowest energy state of the molecule. The
resulting density matrix $\rho = | \psi \rangle\langle \psi |$ has only
one non-zero element:  $\langle i | \rho | j \rangle = \delta_{i 1}
\delta_{j 1}$ ($i,j = 1,2,3,4$).  Thermalization impairs the pure
initial state of (\ref{density2}) and results in a mixed state
$\tilde{\rho}$ of the form
\begin{equation}
\label{rho-thermal}
\tilde{\rho} = \left( \begin{array}{cccc} 1 - \Delta & 0 & 0 & 0 \\ 0 &
\Delta & 0 & 0 \\ 0 & 0 & 0 & 0 \\ 0 & 0 & 0 & 0 \\
\end{array} \right) ,
\end{equation}
where $\Delta = \exp( -E_{v} / k_{B} T )$, $k_{B}$ is the Boltzmann
constant, $T$ is the temperature, and $E_{v}$ is the energy separation
between the vibrational levels on the ground electronic surface.
Thermal excitations to all energy levels higher than $|n_{v} = 1
\rangle$ are neglected, which is a reasonable approximation at
temperatures of about 100~K and below.

\begin{figure}[bt]
\epsfxsize=0.85\textwidth \centerline{\epsffile{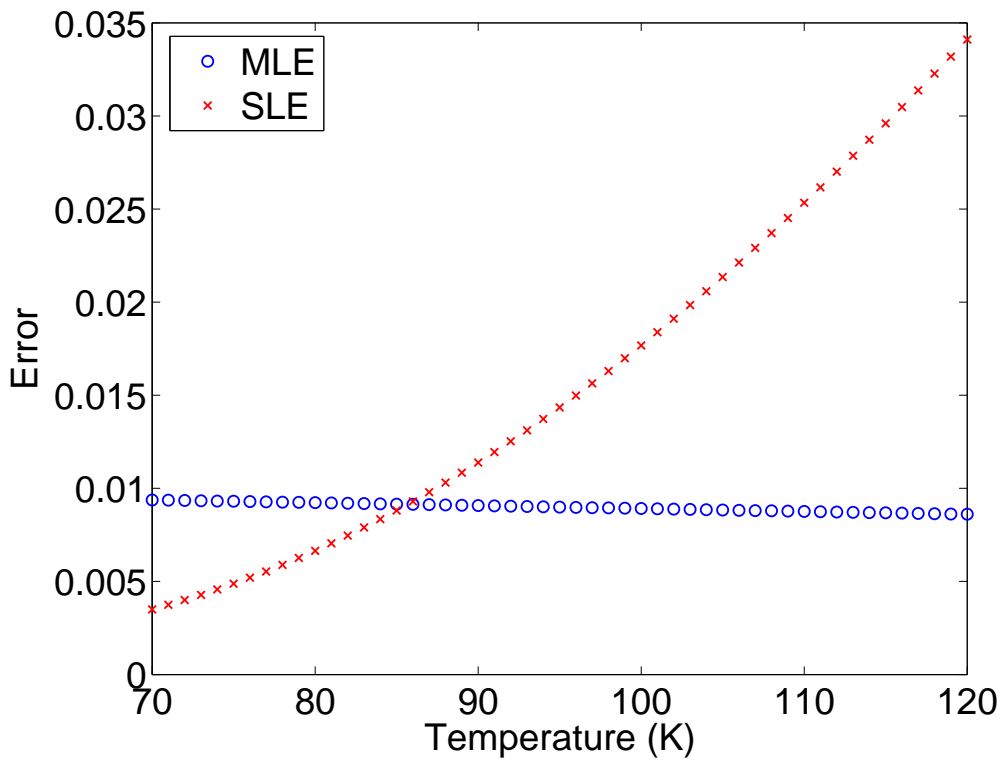}}
\caption{Errors of MLE and SLE bit-flip operations, $\varepsilon
[\rho,\tilde{\rho}]$ of (\ref{error-MLE}) and $\varepsilon'
[\rho,\tilde{\rho}]$ of (\ref{error-SLE}), respectively, for the
thermalized initial state $\tilde{\rho}$ of (\ref{rho-thermal}), versus
the temperature.}
\label{fig:thermal}
\end{figure}

At zero temperature (i.e., when  $\tilde{\rho} = \rho$), the errors of
MLE and SLE bit-flip operations (as defined in (\ref{error-MLE}) and
(\ref{error-SLE}), respectively) are $\varepsilon = 0.0097$ (for
$U_{x}$) and $\varepsilon' = 0.0009$ (for $U'_{x}$), respectively.  This
is consistent with the fact that optimal controls found by the GA have a
higher fidelity for the SLE operations in comparison with the MLE ones,
as shown in table~\ref{fitness}.  However, the situation is strikingly
different in the presence of thermalization.  We evaluate the errors for
both MLE and SLE operations, $\varepsilon [\rho,\tilde{\rho}]$ and
$\varepsilon' [\rho,\tilde{\rho}]$, respectively, for the range of
temperatures between 70~K and 120~K (corresponding to values of $\Delta$
between approximately 0.037 and 0.129).  These errors are shown in
figure~\ref{fig:thermal} versus the temperature.  The error of the SLE
operation increases quite rapidly with the temperature and at
temperatures above 85~K the performance of the SLE operation is worse
than that of the MLE one. This increase of the SLE operation error is
explained by the fact that at higher temperatures a larger portion of
the population is transferred out of the logical basis.   On the other
hand, the performance of the MLE operation changes very little with
thermalization (in fact, the error of the MLE operations slightly
decreases as the temperature increases). This is explained by the fact
that the population transfer \emph{within} the encoding subspace (caused
by thermalization) does not affect the logical state of a qubit with
MLE, as discussed in section \ref{sec:dense}.  Thus, the pure state
$\rho$ and the thermalized state $\tilde{\rho}$ are equally suitable for
the MLE operation due to their logical equivalence.

\section{Two-Qubit Operations for Multilevel Encoding}
\label{sec:2-qubits}

For a two-qubit system, the formalism of MLE is analogous to that
developed in the previous sections for a single qubit.  As before, the
energy levels of a quantum system realizing the $k$th qubit are divided
into two distinct encoding subspaces of equal dimension,
$\mathcal{S}_{0k}$ and $\mathcal{S}_{1k}$.  Let $|\mathbi{a}\rangle_{k}$
denote \emph{any} element of $\mathcal{S}_{0k}$ such that
${}_{k}\langle\mathbi{a}|\mathbi{a}\rangle_{k} = 1$ and
$|\mathbi{b}\rangle_{k}$ denote \emph{any} element of $\mathcal{S}_{1k}$
such that ${}_{k}\langle\mathbi{b}|\mathbi{b}\rangle_{k} = 1$, i.e.,
\begin{gather}
\label{encode_a2}
\nonumber |\mathbi{a}\rangle_{k} = \sum_{i = 1}^{n} a_{i k}
|\chi_{i}\rangle_{k} = \Big( a_{1k} ~\cdots~ a_{n k} \Big)^{T} \\
\text{such that} \ \sum_{i = 1}^{n} |a_{i k}|^{2} = 1
\end{gather}
and
\begin{gather}
\label{encode_b2}
\nonumber |\mathbi{b}\rangle_{k} = \sum_{i = 1}^{n} b_{i k}
|\phi_{i}\rangle_{k} = \Big( b_{1k} ~\cdots~ b_{n k} \Big)^{T} \\
\text{such that} \ \sum_{i = 1}^{n} |b_{i k}|^{2} = 1,
\end{gather}
where $k \in \{1,2\}$ denotes qubit 1 or qubit 2, $n$ is the dimension
of \emph{all} encoding subspaces, and $\{ |\chi_{i}\rangle_{k} \}$ and
$\{|\phi_{i}\rangle_{k} \}$ are orthonormal bases that span
$\mathcal{S}_{0k}$ and $\mathcal{S}_{1k}$, respectively.
Correspondingly, ${}_{k}\langle\chi_{i}|\chi_{j}\rangle_{k'} =
{}_{k}\langle\phi_{i}|\phi_{j}\rangle_{k'} = \delta_{i j}\delta_{k k'}$
and ${}_{k}\langle\chi_{i}|\phi_{j}\rangle_{k'} = 0$.  Now the logical
basis of the $k$th qubit is expressed as
\begin{subequations}
\label{def2}
\begin{gather}
|0\rangle_{k} = |0\rangle_{\text{L}} \otimes |\mathbi{a}\rangle_{k} =
 \Big( a_{1k} ~\cdots~ a_{n k} ~\underbrace{0 ~\cdots~ 0}_{n} \Big)^{T}
 \\ |1\rangle_{k} = |1\rangle_{\text{L}} \otimes |\mathbi{b}\rangle_{k}
 = \Big( \underbrace{0 ~\cdots~ 0}_{n} ~b_{1k} ~\cdots~ b_{n k}
 \Big)^{T}.
\end{gather}
\end{subequations}
It follows that ${}_{k}\langle 0|0\rangle_{k'} = {}_{k}\langle
1|1\rangle_{k'} = \delta_{k k'}$ and ${}_{k}\langle 0|1\rangle_{k'} =
0$.  In the abbreviated notation,
\begin{equation}
|0\rangle_{k} = \left( \begin{array}{c} |\mathbi{a}\rangle_{k} \\ 0
\end{array} \right),
\hspace{2.5mm} |1\rangle_{k} = \left( \begin{array}{c} 0 \\
|\mathbi{b}\rangle_{k}
\end{array} \right),
\end{equation}
and the state vector for the $k$th qubit can be written as
\begin{equation}
|\psi\rangle_{k} = c_{0 k} |0\rangle_{k} + c_{1 k} |1\rangle_{k} =
\left( \begin{array}{c} \alpha_{k} \\ \beta_{k}
 \end{array} \right),
\end{equation}
where $| c_{0 k} |^{2} + | c_{1 k} |^{2} = 1$ (to ensure normalization)
and $\alpha_{k} = c_{0 k} |\mathbi{a}\rangle_{k}$ and $\beta_{k} = c_{1
k} |\mathbi{b}\rangle_{k}$.

The logical basis of the two-qubit system can be defined as
\begin{subequations}
\begin{align}
& |0\rangle_{1} \otimes |0\rangle_{2} = \left( \begin{array}{c}
|\mathbi{a}\rangle_{1} \\ 0
\end{array} \right) \otimes
\left( \begin{array}{c} |\mathbi{a}\rangle_{2} \\ 0
\end{array} \right), \\
& |0\rangle_{1} \otimes |1\rangle_{2} = \left( \begin{array}{c}
|\mathbi{a}\rangle_{1} \\ 0
\end{array} \right) \otimes
\left( \begin{array}{c} 0 \\ |\mathbi{b}\rangle_{2}
\end{array} \right), \\
& |1\rangle_{1} \otimes |0\rangle_{2} = \left( \begin{array}{c} 0 \\
|\mathbi{b}\rangle_{1}
\end{array} \right) \otimes
\left( \begin{array}{c} |\mathbi{a}\rangle_{2} \\ 0
\end{array} \right), \\
& |1\rangle_{1} \otimes |1\rangle_{2} = \left( \begin{array}{c} 0 \\
|\mathbi{b}\rangle_{1}
\end{array} \right) \otimes
\left( \begin{array}{c} 0 \\ |\mathbi{b}\rangle_{2}
\end{array} \right).
\end{align}
\end{subequations}
However, it is much more convenient to use another basis, which is
obtained by a unitary transformation on the basis defined above.  The
preferred logical basis for the two-qubit (TQ) system is
\begin{subequations}
\label{basis2q}
\begin{align}
& |1\rangle_{\text{TQ}} = R \left( |0\rangle_{1} \otimes |0\rangle_{2}
\right) = \left( \begin{array}{c} |\mathbi{a}\rangle_{1} \otimes
|\mathbi{a}\rangle_{2} \\ 0 \\ 0 \\ 0
\end{array} \right), \\
& |2\rangle_{\text{TQ}} = R \left( |0\rangle_{1} \otimes |1\rangle_{2}
\right) = \left( \begin{array}{c} 0 \\ |\mathbi{a}\rangle_{1} \otimes
|\mathbi{b}\rangle_{2} \\ 0 \\ 0
\end{array} \right), \\
& |3\rangle_{\text{TQ}} = R \left( |1\rangle_{1} \otimes |0\rangle_{2}
\right) = \left( \begin{array}{c} 0 \\ 0 \\ |\mathbi{b}\rangle_{1}
\otimes |\mathbi{a}\rangle_{2} \\ 0
\end{array} \right), \\
& |4\rangle_{\text{TQ}} = R \left( |1\rangle_{1} \otimes |1\rangle_{2}
\right) = \left( \begin{array}{c} 0 \\ 0 \\ 0 \\ |\mathbi{b}\rangle_{1}
\otimes |\mathbi{b}\rangle_{2}
\end{array} \right),
\end{align}
\end{subequations}
where $R$ is a $4 n^{2} \times 4 n^{2}$ unitary transformation.  Now
consider $|\Psi\rangle$, a state vector of the two-qubit system.  In the
logical basis of (\ref{basis2q}), $|\Psi\rangle$ is expressed as
\begin{equation}
|\Psi\rangle = \sum_{i = 1}^{4} c_{i} |i\rangle_{\text{TQ}},
\end{equation}
where $\sum_{i = 1}^{4} |c_{i}|^{2} = 1$ ensures that
$\langle\Psi|\Psi\rangle = 1$ ($|\Psi\rangle$ is a vector of dimension
$4 n^{2}$).

In order to formulate the principle of logical equivalence for the
two-qubit system, we introduce a set of 15 EA operators,
\begin{equation}
\Lambda_{r} = \lambda_{r} \otimes \openone_{n^{2}}, \ \ r = \{
1,2,\ldots,15 \},
\end{equation}
where $\openone_{n^{2}}$ is the $n^{2} \times n^{2}$ identity matrix and
$\{ \lambda_{r} \}$ is the set of $4 \times 4$ matrices which represent
the 15 generators of SU(4), which, with $\openone_{4}$, span the space
of all $4 \times 4$ matrices.  The principle of logical equivalence is
formulated analogously to the single-qubit case.  Two physical states,
$|\Psi\rangle$  and $|\Psi'\rangle$, are logically equivalent (denoted
as $|\Psi\rangle \sim |\Psi'\rangle$) if
\begin{equation}
\langle\Psi| \Lambda_{r} |\Psi\rangle = \langle \Psi' | \Lambda_{r} |
\Psi'\rangle, \ \ r = \{ 1,2,\ldots,15 \}.
\end{equation}

Now we develop the explicit form of logical operations for the two-qubit
system with MLE of the logical basis states.  First, consider local
operations which act separately on the two qubits.  Such an operation is
represented (in the preferred basis) by the matrix
\begin{equation}
U = R \left( U^{(1)} \otimes U^{(2)}\right) R^{\dagger},
\end{equation}
where $U^{(k)}$ acts on the $k$th qubit.  In particular, the weak
identity operation that maintains logical equivalence for the entire
two-qubit system and simultaneously for each of the individual qubits is
given by
\begin{equation}
\label{UI2}
U_{I} = R \left( U_{I}^{(1)} \otimes U_{I}^{(2)} \right) R^{\dagger} =
\openone_{4} \otimes \left[ V_{0}^{(1)} \otimes V_{0}^{(2)} \right],
\end{equation}
where $U_{I}^{(k)}$ is the weak identity operation and $V_{0}^{(k)}$ is
an arbitrary $n \times n$ unitary matrix acting on the $k$th qubit.  As
in the single-qubit case, the two-qubit weak identity satisfies
\begin{equation}
U_{I} |\Psi\rangle \sim |\Psi\rangle \ \ \text{for any} \ |\Psi\rangle.
\end{equation}

The proof given in (\ref{proof1}) and (\ref{proof2}) easily extends to
the two-qubit case, so any operation $U$ such that
\begin{equation}
U \in \text{U}(4) \otimes \left[ \text{U}(n) \otimes \text{U}(n) \right]
\subset \text{U}(4n^{2})
\end{equation}
satisfies $U_{I} U \sim U U_{I}$.  Therefore, any two-qubit logical
operation $S$ resulting from SLE will be represented in the framework of
MLE by a matrix of the form
\begin{equation}
\label{2QG}
U_{S} = S \otimes \left[ V^{(1)} \otimes V^{(2)} \right],
\end{equation}
where $S \in \text{U}(4)$ and $V^{(k)} \in \text{U}(n)$.  Since
$V^{(1)}$ and $V^{(2)}$ are arbitrary unitary matrices, any logical
operation will be represented by an infinite number of logically
equivalent physical transformations.

The tensor-product form $V^{(1)} \otimes V^{(2)}$ of (\ref{2QG}) is
required in order to ensure that entanglement created by a logical
operation is reversible not only by a specific inverse operation, but by
any member of a class of logically equivalent operations.  To
demonstrate this point, consider the C-NOT gate, $U_{\text{CN}}$.  For
MLE, this operation is represented by the following set of logically
equivalent $(4n^{2}) \times (4n^{2})$ matrices:
\begin{align}
\label{CN-gate}
\nonumber U_{\text{CN}} &= \left(\begin{array}{cccc} 1 & 0 & 0 & 0 \\ 0
& 1 & 0 & 0 \\ 0 & 0 & 0 & 1 \\ 0 & 0 & 1 & 0
\end{array} \right)
\otimes \left[ V^{(1)} \otimes V^{(2)} \right] \\ &=
\left(\begin{array}{cccc} V^{(1)} \otimes V^{(2)} & 0 & 0 & 0 \\ 0 &
V^{(1)} \otimes V^{(2)} & 0 & 0 \\ 0 & 0 & 0 & V^{(1)} \otimes V^{(2)}
\\ 0 & 0 & V^{(1)} \otimes V^{(2)} & 0
\end{array} \right).
\end{align}
Equation (\ref{CN-gate}) defines an entire class of logically equivalent
operations.  With SLE, applying the C-NOT operation twice results in the
identity operation.  Thus, applying the operation $U_{\text{CN}}$ twice
should result in a logically equivalent state, i.e.,
\begin{equation}
U_{\text{CN}}^{2} |\Psi\rangle \sim |\Psi\rangle.
\end{equation}
Indeed it is clear that $U_{\text{CN}}^{2} \sim U_{I}$. Moreover,
applying $U_{\text{CN}}$ twice actually means that we can sequentially
apply two physically different transformations which belong to the same
equivalence class defined by (\ref{CN-gate}) since the TPS assures that
for any pair of logically equivalent operations $U_{1} \in
U_{\text{CN}}$ and $U_{2} \in U_{\text{CN}}$, the overall operation
$U_{1} U_{2}$ is equivalent to the weak identity $U_{I}$ of (\ref{UI2}).

The C-NOT gate of (\ref{CN-gate}), together with the Hadamard gate of
(\ref{H-gate}) and phase-shift gate of (\ref{P-gate}), establish
universality \cite{DiV95} for quantum computation with MLE.

\section{Multi-qubit Operations for Multilevel Encoding}
\label{sec:m-qubits}

In the case of $M$ qubits, logical equivalence is based on the set of
$Z_{M} = (2^{2 M} - 1)$ EA operators
\begin{equation}
\Lambda_{r} = \lambda_{r} \otimes \openone_{n^{M}}, \ \ r = \{ 1, 2,
\ldots, Z_{M} \}
\end{equation}
where $\openone_{n^{M}}$ is the $n^{M} \times n^{M}$ identity matrix and
$\{ \lambda_{r} \}$ is the set of $2^{M} \times 2^{M}$ matrices which
represent the $Z_{M}$ generators of SU($2^{M}$).  A possible choice of
these generators is, for example, the standard Cartan-Weyl basis or its
Hermitian variant \cite{BaRa86}.  The principle of logical equivalence
for a pair of multi-qubit states $|\Psi\rangle$ and $|\Psi'\rangle$ will
be expressed as usual:
\begin{equation}
|\Psi\rangle \sim |\Psi'\rangle \Leftrightarrow \langle\Psi| \Lambda_{r}
|\Psi\rangle = \langle \Psi' | \Lambda_{r} | \Psi'\rangle \ \ \text{for 
all} \ r.  
\end{equation}
In addition, the states must be normalized.

Unitary logical operations for $M$ qubits, with $n$-dimensional encoding
of every logical basis state, will be elements of the group
\begin{equation}
\text{U}(2^{M}) \otimes \underbrace{ \left[ \text{U}(n) \otimes \cdots
\otimes \text{U}(n) \right] }_{M~\text{times}} \subset \text{U} \left(
(2 n)^{M} \right).
\end{equation}
Table~\ref{t-summary} summarizes the group-theoretic structure of this
mapping from single-level to multilevel encoding.

\begin{table}
\caption{The group-theoretic structure of unitary logical
operations for single-level and multilevel encoding of qubits.}
\label{t-summary}
\begin{indented}
\item[]\begin{tabular}{@{}cccc} \br Encoding & Single qubit & Two qubits
& $M$ qubits \\ \mr $ n = 1 $ & U(2) & U(4) & U($2^{M}$) \\ $ n > 1 $ &
$\text{U}(2) \otimes \text{U}(n)$ & $\text{U}(4) \otimes \left[
\text{U}(n) \otimes \text{U}(n) \right]$ & $\text{U}(2^{M}) \otimes
\underbrace{ \left[ \text{U}(n) \otimes \cdots \otimes \text{U}(n)
\right] }_{M~\text{times}}$ \\ \br
\end{tabular}
\end{indented}
\end{table}

The generalization of the MLE formalism to mixed states and non-unitary
logical operations for multiple qubits is straightforward and analogous
to the single-qubit case (see section~\ref{sec:dense}).

\section{Conclusions}
\label{sec:conc}

The formalism of MLE of logical basis states presented in this paper is
motivated primarily by practical realities such as imperfections and
noise in controls, thermally excited initial states, and
environmentally-induced decoherence.

With MLE, a given logical operation is realized by an infinite number of
equivalent physical transformations which may be used interchangeably
throughout the computation, thereby allowing for a significant
flexibility in the laboratory realizations of control.  A crucial point
is that MLE naturally suits the application of ultrafast broadband
controls which simultaneously drive multiple transitions and have the
advantage of generating faster quantum operations, helping to lessen the
effect of decoherence.

As presented in section~\ref{sec:example} with a simplified model of a
single qubit based on $\text{Na}_{2}$ (which is not efficiently scalable
for quantum computing applications in the absence of entanglement), the
simulation of closed-loop optimal control of single-qubit MLE operations
demonstrates that ultrafast optimal solutions for multilevel systems are
readily achievable.  Moreover, these solutions are not affected by
various decohering processes which are detrimental to quantum
information processing based on SLE.

An open question is whether the multiplicity of logically equivalent
physical transformations will facilitate the discovery of controls with
increased robustness against noise in the control fields.  While this
assumption is intuitively plausible, a rigorous analysis of the
robustness is still required to make any formal conclusions.  A powerful
method for the robustness analysis is the study of the landscape for
optimal control solutions \cite{RHR04, RHR05, CB06}.  Other numerical
approaches to the robustness analysis are being considered as well.

An extremely promising feature of MLE is the ability to work with mixed
initial states and decoherence without a loss of operational fidelity.
If mixing and decoherence are contained within the encoding subspaces,
then coherence between the logical basis states will not be affected,
meaning that quantum computation with full fidelity is possible.  This
property, illustrated by a numerical example in
section~\ref{sec:example}, makes MLE a very attractive approach to
practical quantum computation.

The tensor-product structure of the Hilbert space, which appears in MLE,
is a very general feature of multiparticle systems.  Therefore, there
are many possibilities for building an MLE structure in various physical
systems.  Interesting examples of systems which could be suitable for
MLE are molecules and trapped ions (with encoding of the logical basis
into subspaces of vibrational levels of a molecule or motional sidebands
of an ion, respectively).  Although we are planning experiments
involving diatomic molecules such as $\text{Na}_{2}$ and $\text{K}_{2}$
to study the basic functioning of MLE in ultrafast single-qubit
operations in the presence of thermalization, we cannot yet specify
particular quantum information systems which will benefit most from the
use of MLE, since this depends on many technical considerations.
Nevertheless, we estimate that the tendency toward faster quantum
operations will definitely favour the application of MLE.

\ack This work was supported by the NSF, DARPA, and NSERC.

\appendix
\section{Properties of the Density Matrix Components}
\label{app:a}

In (\ref{density}), we expressed the density matrix of an arbitrary
state in the expanded $(2n)$-dimensional Hilbert space as
\begin{equation}
\label{density:a}
\rho = \sum_{i,j = 1}^{2} r_{i j}|i\rangle_\mathrm{L}\,
{}_\mathrm{L}\!\langle j| \otimes R_{i j} = \left( \begin{array}{cc}
r_{11} R_{11} & r_{12} R_{12} \\ r_{21} R_{21} & r_{22} R_{22}
\end{array} \right),
\end{equation}
where $r_{i j}$ are matrix elements representing the state of the
logical component and $R_{i j}$ are $n \times n$ matrices representing
the state of the encoded component.  This is a generalization of a pure
state, (\ref{qubit}):
\begin{equation}
|\psi\rangle = c_{0}|0\rangle_{\text{L}} \otimes |\mathbi{a}\rangle +
c_{1}|1\rangle_{\text{L}} \otimes |\mathbi{b}\rangle.
\end{equation}
If the dimension of the encoding subspaces decreases from $n$ to 1, then
$\rho$ and $|\psi\rangle$ will represent the logical state of a
two-dimensional quantum system:
\begin{equation}
\rho_{\mathrm{L}} = \left( \begin{array}{cc} r_{11} & r_{12} \\ r_{21} &
r_{22}
\end{array} \right)
\end{equation}
and
\begin{equation}
|\psi\rangle_{\mathrm{L}} = c_{0}|0\rangle_{\mathrm{L}} +
 c_{1}|1\rangle_{\mathrm{L}}.
\end{equation}
Thus, for encoding subspaces of any dimension, the logical component
($\rho_{\mathrm{L}}$ or $|\psi\rangle_{\mathrm{L}} $) represents the
state of a two-level qubit.  Therefore, $\rho_{\mathrm{L}}$ is a proper
density matrix, i.e, it is Hermitian, positive, and normalized.  This
normalization means that $\mathrm{Tr} \left( \rho_{\mathrm{L}} \right) =
r_{11} + r_{22} = 1$, where $r_{11}, \, r_{22} \geq 0$.

In this appendix, we will also show that the matrices $R_{11}$ and
$R_{22}$ of the encoded component are proper density matrices that
satisfy all necessary properties.  The Hermiticity of $R_{11}$ and
$R_{22}$ follows directly from the Hermiticity of $\rho$.  Examining the
matrix structure of $\rho$ in (\ref{density:a}), consider the following
scenarios:
\begin{enumerate}
\item{If $1 \leq k, l \leq n$, then $\rho_{kl} = (R_{11})_{kl}$, which
    implies that $(R_{11})_{kl} = (R_{11})_{l k}^{*}$, meaning $R_{11} =
    R_{11}^{\dagger}$.}
\item{If $n+1 \leq k, l \leq 2n$, then $\rho_{kl} = (R_{22})_{k' l'}$,
    which implies that $(R_{22})_{k' l'} = (R_{22})_{l' k'}^{*}$,
    meaning $R_{22} = R_{22}^{\dagger}$ (where $k' = k - n$ and $l' = l
    - n$).}
\end{enumerate}
However, if $1 \leq k \leq n$ and $n+1 \leq l \leq 2n$, then $\rho_{kl}
= (R_{12})_{kl'}$, which implies that $(R_{12})_{kl'} = (R_{21})_{k'
l}^{*}$, meaning $R_{12} = R_{21}^{\dagger}$.

The positivity of $R_{11}$ and $R_{22}$ also follows directly from the
positivity of $\rho$:
\begin{equation}
\langle\psi|\rho|\psi\rangle \geq 0 \ \ \text{for all} \ |\psi\rangle,
\end{equation}
where
\begin{align}
\nonumber \langle\psi|\rho|\psi\rangle & =
|c_{0}|^{2}r_{11}\langle\mathbi{a}|R_{11}|\mathbi{a}\rangle +
c_{0}^{*}c_{1}r_{12}\langle\mathbi{a}|R_{12}|\mathbi{b}\rangle \\ & +
c_{0}c_{1}^{*}r_{21}\langle\mathbi{b}|R_{21}|\mathbi{a}\rangle +
|c_{1}|^{2}r_{22}\langle\mathbi{b}|R_{22}|\mathbi{a}\rangle \geq 0.
\end{align}
In particular, if $|\psi\rangle = |0\rangle_{\mathrm{L}} \otimes
|\mathbi{a}\rangle$, then $\langle\psi|\rho|\psi\rangle =
r_{11}\langle\mathbi{a}|R_{11}|\mathbi{a}\rangle \geq 0$, which implies
that $\langle\mathbi{a}|R_{11}|\mathbi{a}\rangle \geq 0$ for any
$|\mathbi{a}\rangle$.  Similarly, if $|\psi\rangle =
|1\rangle_{\mathrm{L}} \otimes |\mathbi{b}\rangle$, then
$\langle\mathbi{b}|R_{22}|\mathbi{b}\rangle \geq 0$ for any
$|\mathbi{b}\rangle$.

Now consider the normalization condition for the density matrix $\rho$:
\begin{equation}
\mathrm{Tr} \left( \rho \right) = r_{11}\mathrm{Tr} \left( R_{11}
\right) + r_{22}\mathrm{Tr} \left( R_{22} \right) = 1.
\end{equation}
Using the normalization of the logical component, $r_{11} + r_{12} = 1$,
we obtain
\begin{equation}
\label{norm-condition}
r_{11}\mathrm{Tr} \left( R_{11} \right) + \left( 1 - r_{11} \right)
\mathrm{Tr} \left( R_{22} \right) = 1,
\end{equation}
Since the logical and encoding components are independent, i.e., any
logical state can be ``attached'' to any encoding configuration,
(\ref{norm-condition}) is satisfied \emph{only} when
\begin{equation}
\mathrm{Tr} \left( R_{11} \right) = \mathrm{Tr} \left( R_{22} \right) =
1.
\end{equation}
However, note that $\text{Tr}\left(R_{12}\right) \neq 1$ and
$\text{Tr}\left(R_{21}\right) \neq 1$ in general.

In conclusion, this appendix demonstrates that the density matrix
components $\rho_{\mathrm{L}}$, $R_{11}$, and $R_{22}$ are Hermitian,
positive, and normalized (of unit trace), and therefore are proper
density matrices.


\end{document}